\documentclass[12pt]{article}
\usepackage{amssymb}
\usepackage[all,2cell,ps]{xy}
\usepackage{verbatim}
\usepackage{epsfig}
\def\e{{\mathbf e}}

\makeatletter
\@addtoreset{equation}{section}

\newtheorem{theorem}{Theorem}[section]
\newtheorem{examp}{Example}[section]
\newtheorem{coroll}{Corollary}[section]
\newtheorem{examps}{Examples}[section]

\newtheorem{lemma}{Lemma}[section]
\newtheorem{remark}{Remark}[section]

\newtheorem{remarks}[remark]{Remarks}
\newtheorem{proposition}{Proposition}[section]
\newtheorem{definition}{Definition}[section]

\def\m{\mathop}
 \def\tr{{\rm Tr}}

\def\br{\begin{remark}\rm\small}
\def\1{{\bf 1}}
\def\er{\end{remark}}
\def\bt{\begin{theorem}\rm}
\def\et{\end{theorem}}
\def\bc{\begin{coroll}\rm}
\def\ec{\end{coroll}}
\def\brs{\begin{remarks}.\\ \rm\small\begin{enumerate}}
\def\ers{\end{enumerate}\end{remarks}}
\def\bx{\begin{examp}\small}
\def\ex{\end{examp}}
\def\bl{\begin{lemma}\small}
\def\el{\end{lemma}}
\def\bxs{\begin{examps}. \rm\begin{enumerate}}
\def\exs{\end{enumerate}\end{examps}}
\def\bd{\begin{definition}}
\def\ed{\end{definition}}
\def\bp{\begin{proposition}\rm}
\def\ep{\end{proposition}}
\def\be{\begin{equation}}
\def\ee{\end{equation}}

\def\bea{\begin{eqnarray}}
\def\eea{\end{eqnarray}}
\def\beas{\begin{eqnarray*}}
\def\eeas{\end{eqnarray*}}

\def\C{{\mathbb C}}

\def\R{{\mathbb R}}

\def\ovl{\overline}

\def\eq{eq.\ref}

\def\td{\tilde}

\def\virg{\,\, , \,\,\,\,\,}

\newcommand\encadremath[1]{\vbox{\hrule\hbox{\vrule\kern8pt
\vbox{\kern8pt \hbox{$\displaystyle #1$}\kern8pt}
\kern8pt\vrule}\hrule}}
\def\enca#1{\vbox{\hrule\hbox{
\vrule\kern8pt\vbox{\kern8pt \hbox{$\displaystyle #1$}
\kern8pt} \kern8pt\vrule}\hrule}}

\begin{document}

\sloppy


\pagestyle{empty}
\hfill SPhT-T05/038
\addtolength{\baselineskip}{0.20\baselineskip}
\begin{center}
\vspace{26pt}
{\large \bf {Loop  equations for the semiclassical 2-matrix model with hard edges}}
\newline
\vspace{26pt}

{\sl B.\ Eynard}\hspace*{0.05cm}\footnote{ E-mail: eynard@cea.fr }\\
\vspace{6pt}
Service de Physique Th\'{e}orique de Saclay,\\
F-91191 Gif-sur-Yvette Cedex, France.\\
\end{center}

\vspace{20pt}
\begin{center}
{\bf Abstract}:
The 2-matrix model can be defined in a setting more general than polynomial potentials, namely,
the semiclassical matrix model. In this case, the potentials are such that their derivatives are rational functions,
and the integration paths for eigenvalues are arbitrary homology classes of paths for which the integral is convergent.
This choice includes in particular the case where the integration path has fixed endpoints, called hard edges.
The hard edges induce boundary contributions in the loop equations.
The purpose of this article is to give the loop equations in that semicassical setting.

\end{center}

\section{Introduction}

Orthogonal polynomials and biorthogonal polynomials, in the context of random matrices, have been mostly studied for polynomial potentials, on the real axis, or sometimes
on homology classes of contours going from $\infty$ to $\infty$ \cite{Mehta}.
However, it is possible to define matrix models corresponding to a more general context,
in particular, the ``semi-classical'' (called so, because it contains all the classical polynomials).
It is defined as follows (see Bertola \cite{Marcopaths}):

Consider two potentials, (i.e. functions of a complex variable), $V_1(x)$ and $V_2(y)$ whose derivatives $V'_1(x)$ and $V'_2(y)$ are rational functions
(notice that $\infty$ maybe a pole of $V'_1$ (resp. $V'_2$), this is the case if $V'_1$ (resp. $V'_2$) is a polynomial).

Then consider a generalized integration path $\Gamma=\sum_{i,j} \kappa_{i,j} \gamma_i \times \td\gamma_i$, such that the following integral is absolutely convergent:
\be
\int_\Gamma \e^{-{N\over t}[V_1(x)+V_2(y)-xy]} dx dy
\ee
The possible paths $\gamma$ (resp. $\td\gamma$) are described in \cite{Marcopaths}, they can be closed or open:

- If $\gamma$ (resp. $\td\gamma$) is a closed contour, the result is non-zero only if it encloses a singularity of $\e^{-{N\over t}V_1}$ (resp. $\e^{-{N\over t}V_2}$).

- if $\gamma$ (resp. $\td\gamma$) is an open contour, its extremities can be:
\begin{itemize}
\item any point in the complex plane, except at the poles of $V'_1$ (resp. $V'_2$),
\item a simple pole of ${N\over t}V'_1$ (resp. ${N\over t}V'_2$), with residue $\in \R_-$,
\item a degree $\geq 2$ pole of $V'_1$ (resp. $V'_2$), if $\gamma$ (resp. $\td\gamma$) approaches the pole in a sector where $\Re {N\over t}V_1>0$ (resp. $\Re {N\over t}V_2>0$).
\end{itemize}

Some of the extremities are such that $\e^{-{N\over t}[V_1(x)+V_2(y)-xy]}$ vanishes, they are the poles of $V'_1$ (resp. $V'_2$) of degree at least $2$, as well as the simple poles of ${N\over t}V'_1$ (resp. ${N\over t}V'_2$), with residue $\in \R_-$, we will call them ``poles''.
And some extremities are such that $\e^{-{N\over t}[V_1(x)+V_2(y)-xy]}$ does not vanish, they are arbitrary points in the complex plane, we call them ``hard edges'', and write them:
\be
(X_1,X_2,\dots,X_{K_1}) \qquad ({\rm resp.}\,\, (Y_1,Y_2,\dots,Y_{K_2}))
\ee

\bd
Consider a path $\gamma$ in the complex plane.
Define the set of ``normal matrices'' on the path $\gamma$:
\be
H_N(\gamma):=\left\{M\in GL_N(\C) \,\, \backslash \, \exists U\in U(N), \exists (x_1,\dots,x_N)\in \gamma^N, \, M=U{\rm diag}(x_1,\dots,x_N) U^\dagger \right\}
\ee
with a measure:
\be
dM:= \left(\prod_{i>j} (x_i-x_j)^2\right)\,dU\,\prod_{i=1}^N dx_i
\ee
where $dU$ is the Haar measure on $U(N)$, and $dx_i$ are the tangent vectors to the path $\gamma$.
For a given $M\in H_N(\Gamma)$, the matrix $U$ and the eigenvalues $x_i$ are not uniquely defined, they are defined up to a permutation, and a conjugation by a diagonal unitary matrix,
however $dM$ is invariant under these operations, and thus is well defined.
\ed

For example, if $\gamma$ is the real axis we recognize $H_N(\gamma)=H_N$ the set of hermitean matrices.

Then, as in \cite{eynhabilit}, we define the set of normal matrices on a generalized path $\Gamma=\sum_{i,j} \kappa_{i,j} \gamma_i\times \td\gamma_j$:
\bd
Let $\Gamma=\sum_{i,j} \kappa_{i,j} \gamma_i\times \td\gamma_j$ be a linear combination of products of paths.
We define:
\bea
H^2_N(\Gamma)
&:=&\left\{M_1,M_2\in GL_N(\C)2 \,\, \backslash \, \exists U_1,U_2\in U(N)^2, \exists (x_1,\dots,x_N)\in \C^N, \exists (y_1,\dots,y_N)\in \C^N, \right.\cr
&& M_1=U_1{\rm diag}(x_1,\dots,x_N) U_1^\dagger, \,\, M_2=U_2{\rm diag}(y_1,\dots,y_N) U_2^\dagger,\cr
&& \left.\, \,\, \forall\, k=1,\dots, N,\, \exists i_k\,\, x_k\in \gamma_{i_k},\,\, \forall\, k=1,\dots, N,\, \exists j_k\,\, y_k\in \td\gamma_{j_k}\right\}
\eea
with a measure $dM_1\, dM_2$ such that if $(x_1,\dots, x_N)$ are the eigenvalues of $M_1$ and  $(y_1,\dots, y_N)$ are the eigenvalues of $M_2$,
and $x_k\in \gamma_{i_k}$ and $y_k\in\td\gamma_{j_k}$:
\be
dM_1\, dM_2:= \mathop{\det}_{k,l}{\left(\kappa_{i_k,j_l}\e^{x_{i_k}y_{j_l}}\right)}\,\left(\prod_{i>j} (x_i-x_j)\right)\,\left(\prod_{i>j} (y_i-y_j)\right)\,dU_1\,dU_2\,\prod_{i=1}^N dx_i\,\prod_{i=1}^N dy_i
\ee
\ed

Our goal is to compute the Schwinger-Dyson equations, also called loop equations in the context of random matrices, of the following matrix integral:
\be\label{Zdef}
Z:=\int_{H_N^2(\Gamma)} \e^{-{N\over t}\tr[V_1(M_1)+V_2(M_2)-M_1 M_2]}\, dM_1\,dM_2
\ee

\subsection{Notations}

Define the following polynomials, which vanish at all the hard edges:
\be
s(x):=\prod_{j=1}^{K_1} (x-X_j)=\sum_{r=0}^{K_1} s_r x^r
\ee
\be
\td{s}(y):=\prod_{j=1}^{K_2} (y-Y_j)=\sum_{r=0}^{K_2} \td{s}_r y^r
\ee

\medskip

The resolvent, (which is a formal series in its large $x$ expansion), is  defined by:
\be
W(x):={t\over N}\left< \tr {1\over x-M_1}\right>
\ee
And, up to a shift by the potential, it is more convenient to use $Y(x)$:
\be
Y(x):= V'_1(x)-W(x)
\ee

Then, for technical intermediate calculations, define the following expectation values, which are formal series in their large $x$ expansion:
\bea
U(x,y):= {t\over N}\left<\tr {1\over x-M_1}\,{V'_2(y)-V'_2(M_2)\over y-M_2}\right> \cr
\eea
\bea
U(x,y,x'):= \left<\tr {1\over x-M_1}\,{V'_2(y)-V'_2(M_2)\over y-M_2}\,\tr {1\over x'-M_1}\right>_{\rm c} \cr
\eea
\bea
P(x,y):= {t\over N}\left<\tr {V'_1(x)-V'_1(M_1)\over x-M_1}\,{V'_2(y)-V'_2(M_2)\over y-M_2}\right> \cr
\eea
\bea
A(x):= {t\over N}\left<\tr {s(M_1)\over x-M_1}\,\td{s}(M_2)V'_2(M_2)\right> \cr
\eea
\bea
B(x):= {t\over N}\left<\tr {s(M_1)\over x-M_1}\,\td{s}(M_2)\right> \cr
\eea
\bea
B_k(x):= {t\over N}\left<\tr {s(M_1)\over x-M_1}\,M_2^k\right> \cr
\eea
\bea
D(x):=xB(x)+  \sum_{s=0}^{K_2}\sum_{j=0}^{s-1} \td{s}_s\, {t\over N}\left< \tr\, M_2^{s-1-j} \right>\,B_j(x)
\eea

Notice that $U(x,y)$ and $U(x,y,x')$ are rational functions of $y$, and $P(x,y)$, is a  rational function of both $x$ and $y$, whose
poles are the same as the poles of $V'_1$ and $V'_2$.

\section{The loop equations}


Loop equations, i.e. Schwinger-Dyson equations, merely express the fact that an integral is invariant under a change of variable.
Schwinger equations have been extensively studied in the context of field theory, and have been of great improtance for the study of matrix models \cite{ZJDFG}.
For the polynomial 2-matrix model, loop equations were first announced to give algebraic equations by M. Staudacher \cite{staudacher},
and computed precisely for general potentials, including $1/N^2$ corrections by \cite{eynchain, eynchaint, eynmultimat, eynm2m}.
They have been solved recentely by \cite{EOloop2mat} following the method of \cite{eynloop1mat}.
Here, we follow the method of \cite{eynmultimat}.

\subsection{Generalities}

\subsubsection{Changes of variables}

Consider a change of variable in $H_n^2(\Gamma)$, of the form:
\be\label{chgtvargen}
M_1\to M'_1=M_1+\epsilon f(M_1,M_2)+\overline{\epsilon} f^\dagger(M_1,M_2) + O(|\epsilon|^2)
\ee
\br
Under such a change of variable, the eigenvalues of $M'_1$ are no longer on paths $\gamma_{i_k}$, but on some paths $\gamma'_{i_k}$, which, for $\epsilon$ small enough,
are small deformations of $\gamma_{i_k}$, and thus the integral is indeed unchanged.
This is true only if we can deform the contours, i.e. not at the extremities. At poles, the integrand vanishes, so that the integral is still invariant.
At hard edges, the integral is invariant only if $M'_1=M_1$, i.e. $f(M_1,M_2)$ vanishes at hard edges.
\er

\bigskip

To order 1 in $|\epsilon|$, one has:
\bea
&&\exp{\left[-{N\over t}{\rm Tr}\, (V_1(M'_1)+V_2(M_2)-M'_1 M_2)\right]} \cr
&&= \left( 1-{\epsilon}\,{\cal S}(f)-{\overline\epsilon}\,\overline{\cal S}(f) +O(|\epsilon|^2)\right)\,
\exp{\left[-{N\over t}{\rm Tr}\, (V_1(M_1)+V_2(M_2)-M_1 M_2)\right]}
\eea
with
\be
{\cal S}(f) ={N\over t}\,{\rm Tr}\, (V'_1(M_1)-M_2)f(M_1,M_2)
\ee

The measure $dM_1$ is multiplied by a Jacobian $J$, which we expand to order $1$ in $|\epsilon|$:
\be
dM'_1 = J(f)\, dM_1
= (1+\epsilon {\cal J}(f)+\overline\epsilon \overline{\cal J}(f)+O(\epsilon^2)) dM_1
\ee

Loop equations are obtained by writing that the integral is unchanged, i.e.:
\bea\label{defZ}
Z&:=&
\int {\rm d}M_1\,{\rm d}M_2\,\,\exp{\left[-{N\over t}{\rm Tr}\, (V_1(M_1)+V_2(M_2)-M_1 M_2)\right]}\cr
&=&\int {\rm d}M'_1\,{\rm d}M_2\,\,\exp{\left[-{N\over t}{\rm Tr}\, (V_1(M'_1)+V_2(M_2)-M'_1 M_2)\right]}\cr
&=&\int {\rm d}M_1\,{\rm d}M_2\,(1+{\epsilon}\,({\cal J}(f)-{\cal S}(f))+{\overline\epsilon}\,(\overline{\cal J}(f)-\overline{\cal S}(f))+O(|\epsilon|^2)) \cr
&& \qquad \qquad\,\exp{\left[-{N\over t}{\rm Tr}\, (V_1(M_1)+V_2(M_2)-M_1 M_2)\right]}\cr
\eea
Since this must be true for all argument of $\epsilon$, one must have:
\be
0=\int {\rm d}M_1\,{\rm d}M_2\,({\cal J}(f)-{\cal S}(f))
\,\exp{\left[-{N\over t}{\rm Tr}\, (V_1(M_1)+V_2(M_2)-M_1 M_2)\right]}
\ee
i.e.
\be\label{generalloopeq}
\left<{\cal J}(f)\right>=\left<{\cal S}(f)\right>
\ee
This equation is called a loop equation.

Let us emphasize again that \eq{defZ}, and thus \eq{generalloopeq} hold only if $f$ vanishes at the hard edges.

\medskip

\subsubsection{Split and merge rules}

The rules to compute ${\cal J}(f)$ are called split and merge rules and can be found in the
litterature, for instance in \cite{eynmultimat}.
Notice that ${\cal J}$ is linear, and if $f$ is a product, ${\cal J}(f)$ can be computed by the chain rule.
Thus, it is useful to determine ${\cal J}(f)$ for some particular $f$.
For any two matrices $A$ and $B$, one has:

$\bullet$ {\em Split rule}:
\be\label{splitrule}
{\cal J}\left(A{1\over x-M_1}B\right) = \tr A{1\over x-M_1}\,\tr{1\over x-M_1}B
\ee
or equivalentely:
\be\label{splitruled}
{\cal J}\left(A M_1^k B\right) = \sum_{j=0}^{k-1} \tr A M_1^j \,\tr M_1^{k-1-j}B
\ee

$\bullet$ {\em Merge rule}:
\be\label{mergerule}
{\cal J}\left(A\,\tr\left({1\over x-M_1}\,B\right)\right) = \tr {1\over x-M_1}\,A\,{1\over x-M_1}\,B
\ee
or equivalentely:
\be\label{mergeruled}
{\cal J}\left(A\,\tr\left(M_1^k\,B\right)\right) = \sum_{j=0}^{k-1} \tr M_1^j\,A\,M_1^{k-j-1}\,B
\ee

\subsection{Equations}

In this section, we write the change of - variable, i.e. function $f$, and the corresponding loop equation of type \eq{generalloopeq}.
We always write the contribution of the Jacobian ${\cal J}$ in the LHS and the contribution of the action ${\cal S}$ in the RHS.

$\bullet$ computation of $B(x)$:
from
\bea
f(M_1,M_2)
& = & s(M_1){1\over x-M_1}\,{\td{s}(Y(x))-\td{s}(M_2)\over Y(x)-M_2} \cr
& = & s(x){1\over x-M_1}\,{\td{s}(Y(x))-\td{s}(M_2)\over Y(x)-M_2}
- {s(x)-s(M_1)\over x-M_1}\,{\td{s}(Y(x))-\td{s}(M_2)\over Y(x)-M_2}
\eea
(which indeed vanishes at all hard edges) we get:
\bea
&& s(x)W(x){t\over N}\left<\tr {1\over x-M_1}\,{\td{s}(Y(x))-\td{s}(M_2)\over Y(x)-M_2} \right> \cr
&& + s(x){t^2\over N^2}\left<\tr {1\over x-M_1}\,\tr {1\over x-M_1}\,{\td{s}(Y(x))-\td{s}(M_2)\over Y(x)-M_2} \right>_{\rm c} \cr
&& - {t^2\over N^2}\sum_{r=0}^{K_1}\sum_{i=0}^{r-1} s_r \left<\tr M_1^{r-1-i}\,\tr {x^i-M_1^i\over x-M_1}\,{\td{s}(Y(x))-\td{s}(M_2)\over Y(x)-M_2} \right> \cr
&=& {t\over N}\left<\tr {s(M_1)\over x-M_1}\,(V'_1(M_1)-M_2)\,{\td{s}(Y(x))-\td{s}(M_2)\over Y(x)-M_2} \right> \cr
\eea

i.e.
\bea
B(x)
&=&  s(x)\td{s}(Y(x))W(x) \cr
&& - s(x){t\over N}\left<\tr {V'_1(x)-V'_1(M_1)\over x-M_1}\,\,{\td{s}(Y(x))-\td{s}(M_2)\over Y(x)-M_2} \right> \cr
&& -{t\over N}\left<\tr {s(x)-s(M_1)\over x-M_1}\,\td{s}(M_2)\right> \cr
&& -{t\over N}\left<\tr {s(x)-s(M_1)\over x-M_1}\,(V'_1(M_1)-M_2)\,{\td{s}(Y(x))-\td{s}(M_2)\over Y(x)-M_2} \right> \cr
&& + {t^2\over N^2}\sum_{r=0}^{K_1}\sum_{i=0}^{r-1} s_r \left<\tr M_1^{r-1-i}\,\tr {x^i-M_1^i\over x-M_1}\,{\td{s}(Y(x))-\td{s}(M_2)\over Y(x)-M_2} \right> \cr
&& - s(x){t^2\over N^2}\left<\tr {1\over x-M_1}\,\tr {1\over x-M_1}\,{\td{s}(Y(x))-\td{s}(M_2)\over Y(x)-M_2} \right>_{\rm c} \cr
\eea

$\bullet$ computation of $B_k(x)$:
from
\bea
f(M_1,M_2)
& = & s(M_1){1\over x-M_1}\,{Y(x)^k-M_2^k\over Y(x)-M_2} \cr
& = & s(x){1\over x-M_1}\,{Y(x)^k-M_2^k\over Y(x)-M_2}
- {s(x)-s(M_1)\over x-M_1}\,{Y(x)^k-M_2^k\over Y(x)-M_2}
\eea
(which indeed vanishes at all hard edges) we get:
\bea
&& s(x)W(x){t\over N}\left<\tr {1\over x-M_1}\,{Y(x)^k-M_2^k\over Y(x)-M_2} \right> \cr
&& + s(x){t^2\over N^2}\left<\tr {1\over x-M_1}\,\tr {1\over x-M_1}\,{Y(x)^k-M_2^k\over Y(x)-M_2} \right>_{\rm c} \cr
&& - {t^2\over N^2}\sum_{r=0}^{K_1}\sum_{i=0}^{r-1} s_r \left<\tr M_1^{r-1-i}\,\tr {x^i-M_1^i\over x-M_1}\,{Y(x)^k-M_2^k\over Y(x)-M_2} \right> \cr
& = & {t\over N}\left<\tr {s(M_1)\over x-M_1}\,(V'_1(M_1)-M_2)\,{Y(x)^k-M_2^k\over Y(x)-M_2} \right> \cr
\eea

i.e.
\bea
 B_k(x)
&=&  \,Y(x)^k s(x)W(x) \cr
&& -s(x){t\over N}\left<\tr {V'_1(x)-V'_1(M_1)\over x-M_1}\,\,{Y(x)^k-M_2^k\over Y(x)-M_2} \right> \cr
&& -{t\over N}\left<\tr {s(x)-s(M_1)\over x-M_1}\,(V'_1(M_1)-M_2)\,{Y(x)^k-M_2^k\over Y(x)-M_2} \right> \cr
&& -{t\over N}\left<\tr {s(x)-s(M_1)\over x-M_1}\,M_2^k \right> \cr
&& + {t^2\over N^2}\sum_{r=0}^{K_1}\sum_{i=0}^{r-1} s_r \left<\tr M_1^{r-1-i}\,\tr {x^i-M_1^i\over x-M_1}\,{Y(x)^k-M_2^k\over Y(x)-M_2} \right> \cr
&& - s(x){t^2\over N^2}\left<\tr {1\over x-M_1}\,\tr {1\over x-M_1}\,{Y(x)^k-M_2^k\over Y(x)-M_2} \right>_{\rm c} \cr
\eea

$\bullet$ Computation of $D(x)$:

\bea
D(x)
&=& xB(x) \cr
&& +  \sum_{s=0}^{K_2}\sum_{j=0}^{s-1} \td{s}_s\, {t\over N}\left< \tr\, M_2^{s-1-j} \right>\,B_j(x) \cr
&=& x s(x)\td{s}(Y(x))W(x) \cr
&& + s(x)W(x)  {t\over N}\left< \tr\, {\td{s}(Y(x))-\td{s}(M_2)\over Y(x)-M_2} \right>\,    \cr
&& - xs(x){t\over N}\left<\tr {V'_1(x)-V'_1(M_1)\over x-M_1}\,\,{\td{s}(Y(x))-\td{s}(M_2)\over Y(x)-M_2} \right> \cr
&& -x{t\over N}\left<\tr {s(x)-s(M_1)\over x-M_1}\,\td{s}(M_2)\right> \cr
&& -x{t\over N}\left<\tr {s(x)-s(M_1)\over x-M_1}\,(V'_1(M_1)-M_2)\,{\td{s}(Y(x))-\td{s}(M_2)\over Y(x)-M_2} \right> \cr
&& +x {t^2\over N^2}\sum_{r=0}^{K_1}\sum_{i=0}^{r-1} s_r \left<\tr M_1^{r-1-i}\,\tr {x^i-M_1^i\over x-M_1}\,{\td{s}(Y(x))-\td{s}(M_2)\over Y(x)-M_2} \right> \cr
&& - s(x) \sum_{s=0}^{K_2}\sum_{j=0}^{s-1} \td{s}_s\, {t\over N}\left< \tr\, M_2^{s-1-j} \right>\,{t\over N}\left<\tr {V'_1(x)-V'_1(M_1)\over x-M_1}\,\,{Y(x)^j-M_2^j\over Y(x)-M_2} \right> \cr
&& -  \sum_{s=0}^{K_2}\sum_{j=0}^{s-1} \td{s}_s\, {t\over N}\left< \tr\, M_2^{s-1-j} \right>\,{t\over N}\left<\tr {s(x)-s(M_1)\over x-M_1}\,(V'_1(M_1)-M_2)\,{Y(x)^j-M_2^j\over Y(x)-M_2} \right> \cr
&& -  {t\over N}\left<  \tr {s(x)-s(M_1)\over x-M_1}\,(V'_2(M_2)-M_1)\td{s}(M_2) \right> \cr
&& +  \sum_{s=0}^{K_2}\sum_{j=0}^{s-1} \td{s}_s\, {t\over N}\left< \tr\, M_2^{s-1-j} \right>\, {t^2\over N^2}\sum_{r=0}^{K_1}\sum_{i=0}^{r-1} s_r \left<\tr M_1^{r-1-i}\,\tr {x^i-M_1^i\over x-M_1}\,{Y(x)^j-M_2^j\over Y(x)-M_2} \right> \cr
&& - xs(x){t^2\over N^2}\left<\tr {1\over x-M_1}\,\tr {1\over x-M_1}\,{\td{s}(Y(x))-\td{s}(M_2)\over Y(x)-M_2} \right>_{\rm c} \cr
&& - s(x) \sum_{s=0}^{K_2}\sum_{j=0}^{s-1} \td{s}_s\, {t\over N}\left< \tr\, M_2^{s-1-j} \right>\, {t^2\over N^2}\left<\tr {1\over x-M_1}\,\tr {1\over x-M_1}\,{Y(x)^j-M_2^j\over Y(x)-M_2} \right>_{\rm c} \cr
&& +  \sum_{s=0}^{K_2}\sum_{j=0}^{s-1} \td{s}_s\, {t^2\over N^2}\left< \tr\, M_2^{s-1-j} \tr {s(x)-s(M_1)\over x-M_1}\,M_2^j \right>_{\rm c} \cr
\eea

$\bullet$ computation of $A(x)$:
By doing a change of variable on $M_2$ of the form:
\bea
\td{f}(M_1,M_2)
& = & s(M_1){1\over x-M_1}\,\td{s}(M_2) \cr
\eea
we have:
\bea
&& {t^2\over N^2}\sum_{s=0}^{K_2}\sum_{j=0}^{s-1} \td{s}_s \left<\tr\,{s(M_1)\over x-M_1}\,M_2^{j} \, \tr\, M_2^{s-1-j} \right> \cr
&=& A(x) - x B(x) +{t\over N}\left<\tr\,s(M_1)\,\td{s}(M_2)\right> \cr
\eea
i.e.
\bea
A(x)
&=&   D(x) \cr
&& -{t\over N}\left<\tr\,s(M_1)\,\td{s}(M_2)\right> \cr
&& + {t^2\over N^2}\sum_{s=0}^{K_2}\sum_{j=0}^{s-1} \td{s}_s \left<\tr\,{s(M_1)\over x-M_1}\,M_2^{j}\, \tr\, M_2^{s-1-j} \right>_{\rm c} \cr
\eea

$\bullet$ Main computation:
from
\bea
f(M_1,M_2)
& = & s(M_1){1\over x-M_1}\,{\td{s}(Y(x))V'_2(Y(x))-\td{s}(M_2)V'_2(M_2)\over Y(x)-M_2} \cr
& = & s(x){1\over x-M_1}\,{\td{s}(Y(x))V'_2(Y(x))-\td{s}(M_2)V'_2(M_2)\over Y(x)-M_2} \cr
&& - {s(x)-s(M_1)\over x-M_1}\,{\td{s}(Y(x))V'_2(Y(x))-\td{s}(M_2)V'_2(M_2)\over Y(x)-M_2}
\eea
we get:
\bea
&& s(x){t\over N}W(x)\left<\,\tr {1\over x-M_1}\,{\td{s}(Y(x))V'_2(Y(x))-\td{s}(M_2)V'_2(M_2)\over Y(x)-M_2} \right> \cr
&& +s(x){t^2\over N^2}\left<\tr {1\over x-M_1}\,\tr {1\over x-M_1}\,{\td{s}(Y(x))V'_2(Y(x))-\td{s}(M_2)V'_2(M_2)\over Y(x)-M_2} \right>_{\rm c} \cr
&& - {t^2\over N^2}\sum_{r=0}^{K_1}\sum_{i=0}^{r-1} s_r \left<\tr M_1^{r-1-i}\,\tr {x^i-M_1^i\over x-M_1}\,{\td{s}(Y(x))V'_2(Y(x))-\td{s}(M_2)V'_2(M_2)\over Y(x)-M_2} \right> \cr
&=& {t\over N}\left<\tr\, s(M_1)\,{1\over x-M_1}\,(V'_1(M_1)-M_2)\,{\td{s}(Y(x))V'_2(Y(x))-\td{s}(M_2)V'_2(M_2)\over Y(x)-M_2}\right> \cr
\eea
i.e.
\bea
&& s(x){t^2\over N^2}\left<\tr {1\over x-M_1}\,\tr {1\over x-M_1}\,{\td{s}(Y(x))V'_2(Y(x))-\td{s}(M_2)V'_2(M_2)\over Y(x)-M_2} \right>_{\rm c} \cr
&=&  s(x)\td{s}(Y(x))V'_2(Y(x))W(x) \cr
&& - s(x){t\over N}\left<\,\tr {V'_1(x)-V'_1(M_1)\over x-M_1}\,{\td{s}(Y(x))V'_2(Y(x))-\td{s}(M_2)V'_2(M_2)\over Y(x)-M_2} \right> \cr
&&- {t\over N}\left<\tr\, {s(x)-s(M_1)\over x-M_1}\,(V'_1(M_1)-M_2)\,{\td{s}(Y(x))V'_2(Y(x))-\td{s}(M_2)V'_2(M_2)\over Y(x)-M_2}\right> \cr
&& + {t^2\over N^2}\sum_{r=0}^{K_1}\sum_{i=0}^{r-1} s_r \left<\tr M_1^{r-1-i}\,\tr {x^i-M_1^i\over x-M_1}\,{\td{s}(Y(x))V'_2(Y(x))-\td{s}(M_2)V'_2(M_2)\over Y(x)-M_2} \right> \cr
&& - {t\over N}\left<\,\tr {s(x)-s(M_1)\over x-M_1}\,\td{s}(M_2)V'_2(M_2) \right> \cr
&& -A(x) \cr
\eea
i.e.
\bea
&& s(x){t^2\over N^2}\left<\tr {1\over x-M_1}\,\tr {1\over x-M_1}\,{\td{s}(Y(x))V'_2(Y(x))-\td{s}(M_2)V'_2(M_2)\over Y(x)-M_2} \right>_{\rm c} \cr
&& + {t^2\over N^2}\sum_{s=0}^{K_2}\sum_{j=0}^{s-1} \td{s}_s \left<\tr\,{s(M_1)\over x-M_1}\,M_2^{j}\, \tr\, M_2^{s-1-j} \right>_{\rm c} \cr
&=&  s(x)\td{s}(Y(x))V'_2(Y(x))W(x) \cr
&& - s(x){t\over N}\left<\,\tr {V'_1(x)-V'_1(M_1)\over x-M_1}\,{\td{s}(Y(x))V'_2(Y(x))-\td{s}(M_2)V'_2(M_2)\over Y(x)-M_2} \right> \cr
&&- {t\over N}\left<\tr\, {s(x)-s(M_1)\over x-M_1}\,(V'_1(M_1)-M_2)\,{\td{s}(Y(x))V'_2(Y(x))-\td{s}(M_2)V'_2(M_2)\over Y(x)-M_2}\right> \cr
&& + {t^2\over N^2}\sum_{r=0}^{K_1}\sum_{i=0}^{r-1} s_r \left<\tr M_1^{r-1-i}\,\tr {x^i-M_1^i\over x-M_1}\,{\td{s}(Y(x))V'_2(Y(x))-\td{s}(M_2)V'_2(M_2)\over Y(x)-M_2} \right> \cr
&& - {t\over N}\left<\,\tr {s(x)-s(M_1)\over x-M_1}\,\td{s}(M_2)V'_2(M_2) \right> \cr
&& +{t\over N}\left<\tr\,s(M_1)\,\td{s}(M_2)\right> \cr
&& -  D(x) \cr
\eea
Now, inserting the value of $D(x)$, and after tedious but straightforward computations, we get:
\bea
&& s(x)\td{s}(Y(x)){t^2\over N^2}\left<\tr {1\over x-M_1}\,\tr {1\over x-M_1}\,{V'_2(Y(x))-V'_2(M_2)\over Y(x)-M_2} \right>_{\rm c} \cr
&& +s(x){t^2\over N^2}\left<\tr {1\over x-M_1}\,\tr {1\over x-M_1}\,(V'_2(M_2)-M_1)\,{\td{s}(Y(x))-\td{s}(M_2)\over Y(x)-M_2} \right>_{\rm c} \cr
&& - s(x){t^2\over N^2}\left<\tr {1\over x-M_1}\,\tr {\td{s}(Y(x))-\td{s}(M_2)\over Y(x)-M_2} \right>_{\rm c} \cr
&& + s(x) \sum_{s=0}^{K_2}\sum_{j=0}^{s-1} \td{s}_s\, {t^2\over N^2}\left< \tr\, M_2^{s-1-j} \tr {1\over x-M_1}\,M_2^j \right>_{\rm c} \cr
&& - s(x) \sum_{s=0}^{K_2}\sum_{j=0}^{s-1} \td{s}_s\, {t\over N}\left< \tr\, M_2^{s-1-j} \right>\, {t^2\over N^2}\left<\tr {1\over x-M_1}\,\tr {1\over x-M_1}\,{Y(x)^j-M_2^j\over Y(x)-M_2} \right>_{\rm c} \cr
&& +  \sum_{s=0}^{K_2}\sum_{j=0}^{s-1} \td{s}_s\, {t\over N}\left< \tr\, M_2^{s-1-j} \right>\, {t^2\over N^2}\sum_{r=0}^{K_1}\sum_{i=0}^{r-1} s_r \left<\tr M_1^{r-1-i}\,\tr {x^i-M_1^i\over x-M_1}\,{Y(x)^j-M_2^j\over Y(x)-M_2} \right>_{\rm c} \cr
&& - {t^2\over N^2}\sum_{r=0}^{K_1}\sum_{i=0}^{r-1} s_r \left<\tr M_1^{r-1-i}\,\tr {x^i-M_1^i\over x-M_1}\,(V'_2(M_2)-M_1)\,{\td{s}(Y(x))-\td{s}(M_2)\over Y(x)-M_2} \right>_{\rm c} \cr
&=&  s(x)\td{s}(Y(x))\left((V'_2(Y(x))-x)(V'_1(x)-Y(x)) -P(x,y)+t\right) \cr
&& - s(x){t\over N}\left<\,\tr {V'_1(x)-V'_1(M_1)\over x-M_1}\,(V'_2(M_2)-M_1)\,{\td{s}(Y(x))-\td{s}(M_2)\over Y(x)-M_2} \right> \cr
&& + s(x) \sum_{s=0}^{K_2}\sum_{j=0}^{s-1} \td{s}_s\, {t\over N}\left< \tr\, M_2^{s-1-j} \right>\,{t\over N}\left<\tr {V'_1(x)-V'_1(M_1)\over x-M_1}\,\,{Y(x)^j-M_2^j\over Y(x)-M_2} \right> \cr
&&- \td{s}(Y(x)){t\over N}\left<\tr\, {s(x)-s(M_1)\over x-M_1}\,(V'_1(M_1)-M_2)\,{V'_2(Y(x))-V'_2(M_2)\over Y(x)-M_2}\right> \cr
&& + \td{s}(Y(x)){t^2\over N^2}\sum_{r=0}^{K_1}\sum_{i=0}^{r-1} s_r \left<\tr M_1^{r-1-i}\,\tr {x^i-M_1^i\over x-M_1}\,{V'_2(Y(x))-V'_2(M_2)\over Y(x)-M_2} \right> \cr
&&- {t\over N}\left<\tr\, {s(x)-s(M_1)\over x-M_1}\, \right. \cr
&& \qquad\qquad \left. (V'_1(M_1)V'_2(M_2)-M_2V'_2(M_2)-M_1V'_1(M_1)+M_1M_2)\,\,{\td{s}(Y(x))-\td{s}(M_2)\over Y(x)-M_2}\right> \cr
&& - {t^2\over N^2} \left<\tr {s(x)-s(M_1)\over x-M_1}\,\tr \,{\td{s}(Y(x))-\td{s}(M_2)\over Y(x)-M_2} \right> \cr
&& + {t^2\over N^2}\sum_{r=0}^{K_1}\sum_{i=0}^{r-1} s_r \left<\tr M_1^{r-1-i}\right>\,\left<\tr {x^i-M_1^i\over x-M_1}\,(V'_2(M_2)-M_1)\,{\td{s}(Y(x))-\td{s}(M_2)\over Y(x)-M_2} \right> \cr
&& + {t^2\over N^2} \sum_{s=0}^{K_2}\sum_{j=0}^{s-1} \td{s}_s\, \left< \tr\, M_2^{s-1-j} \right>\,\left<\tr {s(x)-s(M_1)\over x-M_1}\,(V'_1(M_1)-M_2)\,{Y(x)^j-M_2^j\over Y(x)-M_2} \right> \cr
&& - {t^3\over N^3} \sum_{r=0}^{K_1}\sum_{i=0}^{r-1} \sum_{s=0}^{K_2}\sum_{j=0}^{s-1} s_r \td{s}_s\,  \left<\tr M_1^{r-1-i}\right>\,\left< \tr\, M_2^{s-1-j} \right>\, \,\left<\tr {x^i-M_1^i\over x-M_1}\,{Y(x)^j-M_2^j\over Y(x)-M_2} \right> \cr
\eea
which we write as follows:
\be\label{maineq}
\encadremath{
s(x)\td{s}(y)\,E(x,Y(x))={t^2\over N^2} L(x)
}\ee
where $E(x,y)$ is a rational fraction of both $x$ and $y$ with poles at the poles of $V'_1$ and $V'_2$, and at the hard edges:
\bea
&& E(x,y) \cr
&=&  (V'_2(y)-x)(V'_1(x)-y) -P(x,y)+t \cr
&& - {1\over \td{s}(y)}{t\over N}\left<\,\tr {V'_1(x)-V'_1(M_1)\over x-M_1}\,(V'_2(M_2)-M_1)\,{\td{s}(y)-\td{s}(M_2)\over y-M_2} \right> \cr
&& + {1\over \td{s}(y)} \sum_{s=0}^{K_2}\sum_{j=0}^{s-1} \td{s}_s\, {t\over N}\left< \tr\, M_2^{s-1-j} \right>\,{t\over N}\left<\tr {V'_1(x)-V'_1(M_1)\over x-M_1}\,\,{y^j-M_2^j\over y-M_2} \right> \cr
&&- {1\over s(x)}{t\over N}\left<\tr\, {s(x)-s(M_1)\over x-M_1}\,(V'_1(M_1)-M_2)\,{V'_2(y)-V'_2(M_2)\over y-M_2}\right> \cr
&& + {1\over s(x)}{t^2\over N^2}\sum_{r=0}^{K_1}\sum_{i=0}^{r-1} s_r \left<\tr M_1^{r-1-i}\,\tr {x^i-M_1^i\over x-M_1}\,{V'_2(y)-V'_2(M_2)\over y-M_2} \right> \cr
&&- {1\over s(x)\td{s}(y)}{t\over N}\left<\tr\, {s(x)-s(M_1)\over x-M_1}\,\right. \cr
&& \qquad \qquad \left.(V'_1(M_1)V'_2(M_2)-M_2V'_2(M_2)-M_1V'_1(M_1)+M_1M_2)\,\,{\td{s}(y)-\td{s}(M_2)\over y-M_2}\right> \cr
&& - {1\over s(x)\td{s}(y)}{t^2\over N^2} \left<\tr {s(x)-s(M_1)\over x-M_1}\,\tr \,{\td{s}(y)-\td{s}(M_2)\over y-M_2} \right> \cr
&& + {1\over s(x)\td{s}(y)}{t^2\over N^2}\sum_{r=0}^{K_1}\sum_{i=0}^{r-1} s_r \left<\tr M_1^{r-1-i}\right>\,\left<\tr {x^i-M_1^i\over x-M_1}\,(V'_2(M_2)-M_1)\,{\td{s}(y)-\td{s}(M_2)\over y-M_2} \right> \cr
&& + {1\over s(x)\td{s}(y)}{t^2\over N^2} \sum_{s=0}^{K_2}\sum_{j=0}^{s-1} \td{s}_s\, \left< \tr\, M_2^{s-1-j} \right>\,\left<\tr {s(x)-s(M_1)\over x-M_1}\,(V'_1(M_1)-M_2)\,{y^j-M_2^j\over y-M_2} \right> \cr
&& - {1\over s(x)\td{s}(y)}{t^3\over N^3} \sum_{r=0}^{K_1}\sum_{i=0}^{r-1} \sum_{s=0}^{K_2}\sum_{j=0}^{s-1} s_r \td{s}_s\,  \left<\tr M_1^{r-1-i}\right>\,\left< \tr\, M_2^{s-1-j} \right>\, \,\left<\tr {x^i-M_1^i\over x-M_1}\,{y^j-M_2^j\over y-M_2} \right> \cr
\eea
and in the right hand side of \eq{maineq}, we have:
\bea
&& L(x) \cr
&=& \left<\tr {1\over x-M_1}\,\tr {1\over x-M_1}\,{V'_2(Y(x))-V'_2(M_2)\over Y(x)-M_2} \right>_{\rm c} \cr
&& +{1\over \td{s}(Y(x))}\left<\tr {1\over x-M_1}\,\tr {1\over x-M_1}\,(V'_2(M_2)-M_1)\,{\td{s}(Y(x))-\td{s}(M_2)\over Y(x)-M_2} \right>_{\rm c} \cr
&& - {1\over \td{s}(Y(x))}\left<\tr {1\over x-M_1}\,\tr {\td{s}(Y(x))-\td{s}(M_2)\over Y(x)-M_2} \right>_{\rm c} \cr
&& + {1\over \td{s}(Y(x))} \sum_{s=0}^{K_2}\sum_{j=0}^{s-1} \td{s}_s\, \left< \tr\, M_2^{s-1-j} \tr {1\over x-M_1}\,M_2^j \right>_{\rm c} \cr
&& - {1\over \td{s}(Y(x))} \sum_{s=0}^{K_2}\sum_{j=0}^{s-1} \td{s}_s\, {t\over N}\left< \tr\, M_2^{s-1-j} \right>\, \left<\tr {1\over x-M_1}\,\tr {1\over x-M_1}\,{Y(x)^j-M_2^j\over Y(x)-M_2} \right>_{\rm c} \cr
&& + {1\over s(x)\td{s}(Y(x))} \sum_{r=0}^{K_1}\sum_{i=0}^{r-1} \sum_{s=0}^{K_2}\sum_{j=0}^{s-1} s_r \td{s}_s\, \cr
&& \qquad\qquad\qquad\qquad\qquad {t\over N}\left< \tr\, M_2^{s-1-j} \right>\,  \left<\tr M_1^{r-1-i}\,\tr {x^i-M_1^i\over x-M_1}\,{Y(x)^j-M_2^j\over Y(x)-M_2} \right>_{\rm c} \cr
&& - {1\over s(x)\td{s}(Y(x))}\sum_{r=0}^{K_1}\sum_{i=0}^{r-1} s_r \left<\tr M_1^{r-1-i}\,\tr {x^i-M_1^i\over x-M_1}\,(V'_2(M_2)-M_1)\,{\td{s}(Y(x))-\td{s}(M_2)\over Y(x)-M_2} \right>_{\rm c} \cr
\eea

\subsection{Examples}

Equation \eq{maineq} looks rather terrible, but it is actualy very simple to use.
Let us illustrate it on simple examples.

\subsubsection{No hard edges}

If there is no hard edges, we have $s(x)=1$ and $\td{s}(y)=1$, and the loop equation becomes:
\be
 (V'_2(Y(x))-x)(V'_1(x)-Y(x)) -P(x,Y(x))+t
= {t^2\over N^2}\left<\tr {1\over x-M_1}\,\tr {1\over x-M_1}\,{V'_2(Y(x))-V'_2(M_2)\over Y(x)-M_2} \right>_{\rm c}
\ee
which is the well known loop equation of the 2-matrix model with polynomial potentials \cite{eynmultimat}.

\subsubsection{1-matrix model}

Consider the 1-matrix model with weight $\e^{-{N\over t}\tr V(M)}$.
It is equivalent to a 2-matrix model where $M_2$ is gaussian, i.e. $V'_2(y)=y$, $\td{s}(y)=1$, $V'(x)=V'_1(x)-x$.
The equation \eq{maineq} becomes:
\bea
E(x,y)
&=&  (y-x)(V'(x)+x-y) -{t\over N}\left<\tr\, {V'(x)-V'(M)\over x-M}\right> \cr
&&- {1\over s(x)}{t\over N}\left<\tr\, {s(x)-s(M)\over x-M}\,V'(M)\right> \cr
&& + {1\over s(x)}{t^2\over N^2}\sum_{r=0}^{K_1}\sum_{i=0}^{r-1} s_r \left<\tr M^{r-1-i}\,\tr {x^i-M^i\over x-M}\, \right> \cr
\eea
and in the right hand side of \eq{maineq}, we have:
\bea
L(x) = \left<\tr {1\over x-M}\,\tr {1\over x-M} \right>_{\rm c}
\eea
Since $E(x,y)$ is quadratic in $y$, this equation defines an hyperelliptical curve.

\subsubsection{1-matrix model with only one hard edge}

In particular, consider $s(x)=(x-a)$, we have:
\be
E(x,y)
=  (y-x)(V'(x)+x-y) -{t\over N}\left<\tr\, {V'(x)-V'(M)\over x-M}\right> - {1\over x-a}{t\over N}\left<\tr\, V'(M)\right>
\ee

\subsubsection{1-matrix model with only two hard edges}

In particular, consider $s(x)=(x-a)(x-b)$, we have:
\bea
E(x,y)
&=&  (y-x)(V'(x)+x-y) -{t\over N}\left<\tr\, {V'(x)-V'(M)\over x-M}\right> \cr
&&- {1\over (x-a)(x-b)}{t\over N}\left<\tr\, (x+M-a-b)\,V'(M)\right>
 + {t^2\over (x-a)(x-b)}   \cr
\eea

\subsubsection{2-matrix model with only one hard edge}

Consider $s(x)=(x-a)$ and $\td{s}(y)=1$, we have:
\bea
E(x,y)
&=&  (V'_2(y)-x)(V'_1(x)-y) -P(x,y)+t \cr
&&- {1\over x-a}{t\over N}\left<\tr\, (V'_1(M_1)-M_2)\,{V'_2(y)-V'_2(M_2)\over y-M_2}\right> \cr
\eea
and in the right hand side of \eq{maineq}, we have:
\bea
L(x)
&=& \left<\tr {1\over x-M_1}\,\tr {1\over x-M_1}\,{V'_2(Y(x))-V'_2(M_2)\over Y(x)-M_2} \right>_{\rm c} \cr
\eea

\subsubsection{2-matrix model with only two hard edges}

Consider $s(x)=(x-a)(x-b)$ and $\td{s}(y)=1$, we have:
\bea
E(x,y)
&=&  (V'_2(y)-x)(V'_1(x)-y) -P(x,y)+t \cr
&&- {1\over s(x)}{t\over N}\left<\tr\, (x+M_1-a-b)\,(V'_1(M_1)-M_2)\,{V'_2(y)-V'_2(M_2)\over y-M_2}\right> \cr
&& + {1\over s(x)}{t^2\over N} \left<\tr {V'_2(y)-V'_2(M_2)\over y-M_2} \right> \cr
\eea
and in the right hand side of \eq{maineq}, we have:
\bea
L(x)
&=& \left<\tr {1\over x-M_1}\,\tr {1\over x-M_1}\,{V'_2(Y(x))-V'_2(M_2)\over Y(x)-M_2} \right>_{\rm c} \cr
\eea

\subsubsection{2-matrix model with only one hard edge in $x$ and one hard edge in $y$}

Consider $s(x)=(x-a)$ and $\td{s}(y)=(y-b)$, we have:
\bea
E(x,y)
&=&  (V'_2(y)-x)(V'_1(x)-y) -P(x,y)+t \cr
&& - {1\over \td{s}(y)}{t\over N}\left<\,\tr {V'_1(x)-V'_1(M_1)\over x-M_1}\,(V'_2(M_2)-M_1)\, \right> \cr
&&- {1\over s(x)}{t\over N}\left<\tr\, (V'_1(M_1)-M_2)\,{V'_2(y)-V'_2(M_2)\over y-M_2}\right> \cr
&& - {1\over s(x)\td{s}(y)}{t\over N}\left<\tr\, (V'_1(M_1)V'_2(M_2)-M_2V'_2(M_2)-M_1V'_1(M_1)+M_1M_2)\,\right> \cr
&& - {t^2\over s(x)\td{s}(y)} \cr
\eea
and in the right hand side of \eq{maineq}, we have:
\bea
L(x)
&=& \left<\tr {1\over x-M_1}\,\tr {1\over x-M_1}\,{V'_2(Y(x))-V'_2(M_2)\over Y(x)-M_2} \right>_{\rm c} \cr
&& +{1\over \td{s}(Y(x))}\left<\tr {1\over x-M_1}\,\tr {1\over x-M_1}\,(V'_2(M_2)-M_1)\, \right>_{\rm c} \cr
\eea

\section{Large $N$ limit, algebraic curve}

In the large $N$ limit, \eq{maineq} reduces to an algebraic equation:
\be
E(x,Y(x))=0
\ee
One should notice that in the large $N$ limit, there is the factorization \cite{ZJDFG}: $<\tr\, \tr>=<\tr><\tr>$, so that $x$ and $y$ play symmetric roles \cite{matytsin}.

We see that $E(x,y)$ has poles only at the poles of $V'_1$ and zeroes of $s$ in $x$
and at the poles of $V'_2$ and the zeroes of $\td{s}$ in $y$.

\subsection{Behaviour near hard edges}

near an hard edge $x\to X_i$, such that $s(X_i)=0$, we have:
\bea
  Y^2(x)
&\m\sim_{x\to X_i}& - {1\over s(x)}\,{t\over N}\left<\tr {s(x)-s(M_1)\over x-M_1}\,(V'_1(M_1)-M_2)\right> \cr
&& + {1\over s(x)}\, \,\sum_{r=2}^{K_1} \sum_{l=1}^{r-1}    s_r {t\over N}\left<\tr M_1^{r-1-l}\right> {t\over N}\left<\tr {x^l-M_1^l\over x-M_1}\right> \cr
&& + {\rm finite}
\eea
Thus:
\bea
\mathop{\rm Res}_{X_i}  Y^2(x) dx
&=& {1\over s'(X_i)}\,{t\over N}\left<\tr {s(M_1)\over X_i-M_1}\,(V'_1(M_1)-M_2)\right> \cr
&& + {1\over s'(X_i)}\, \,\sum_{r=2}^{K_1} \sum_{l=1}^{r-1}    s_r {t\over N}\left<\tr M_1^{r-1-l}\right> {t\over N}\left<\tr {X_i^l-M_1^l\over X_i-M_1}\right> \cr
\eea

Hard edges are at the same time poles of $Y(x)$, and zeroes of $dx$, so that $Y(x)dx$ is finite.

\subsection{Behaviour near poles of the potential}

Near a finite pole $\xi$ of $V'_1(x)$, we have:
\be
Y(x) \m\sim_{x\to\xi} V'_1(x) - W(\xi) + O(x-\xi)
\ee
thus
\be
\mathop{\rm Res}_{\xi}  Y(x) dx = \mathop{\rm Res}_{\xi}  V'_1(x) dx
\ee

\medskip

Near a pole at $\infty$, we have:
\be
Y(x) \m\sim_{x\to\infty} V'_1(x) - {t\over x} + O(1/x^2)
\ee
thus
\be
\mathop{\rm Res}_{\xi}  Y(x) dx = t
\ee

\subsection{Determination of the algebraic equation $E(x,y)$}

So far, we know that $E(x,y)$ is a rational function of $x$ and $y$,
we know its form, and its behaviour near poles,
but most of the coefficients are not determined by the loop equations.

The remaining coefficients of $E(x,y)$, as usual, are determined by extra requirements,
which depend on how the matrix model is defined, i.e. on the purpose for which we introduce the matrix model.

The two most frequent definitions of the matrix model are (we mainly follow the presentation of \cite{eynhabilit}):

\subsubsection{Case of the convergent matrix model}

In this case, the matrix model is defined by the convergent integral \eq{Zdef}.
For generic potentials and hard edges, ${1\over N^2}\ln{Z}$ has a large $N$ limit $F_0$, but has no $1/N$ series expansion.
The resolvent, and thus the function $Y(x)$ also, has a large $N$ limit, but no $1/N$ expansion.
This fact can be understood from the work of \cite{BDE}.

The large $N$ limit of $Y(x)$ obeys an algebraic equation $E(x,Y(x))=0$.

Now, consider an arbitrary $\td{E}(x,y)$ satisfying the correct behaviours near poles.
It gives a function $\td{Y}(x)$, and from it one can compute the free energy $F_0$ (see the formula in \cite{MarcoF}).
The $Y(x)$ which is the large $N$ limit of the resolvent is the one for which $\Re(F_0)$ is minimal (see \cite{eynhabilit,BDE}).
That implies in particular that for any contour ${\cal C}$ on the algebraic curve$E(x,y)=0$, one has:
\be\label{Reointydx}
\forall {\cal C} \,\,\, , \qquad \Re \oint_{{\cal C}} Y(x) dx =0
\ee
On an algebraic curve of the type $E(x,y)=0$, there are at most $2\times{\rm genus}$ independent irreducible cycles, (plus contours around poles),
and one can check that the number of constraints of type \eq{Reointydx}, exactly matches the number of coefficients of $E(x,y)$ not determined by the pole behaviours.
Thus, condition \eq{Reointydx}, determines $E(x,y)$ completely.
In case there are several solutions, one determines a unique one, by choosing the absolute minimum of $F_0$.

\subsubsection{Case of the formal matrix model}

The formal matrix model can be defined in a combinatoric way, as a formal series, generating discrete surfaces.
The formal expansion is obtained by expanding the matrix integral \eq{Zdef} with the Feynman graph technics \cite{BIPZ, Kazakov, ZJDFG}.

In that model, $\ln{Z}$ has a $1/N$ expansion by definition, as well as the resolvent, and all expectation values.
The resolvent of the formal matrix model, is thus obtained by solving the loop equation \eq{maineq}, order by order in $1/N^2$.

\smallskip

That model, in addition to the potentials and hard edges, depends on a ``vacuum'' around which the Feynman expansion is performed.
This vacuum is characterized by a set of ``filling fractions'', as follows (see \cite{eynhabilit} for more details):
The potential $V_1(x)+V_2(y)-xy$ has a certain number of extrema, which are given by the algebraix equation:
\be
V'_1(x)=y \virg V'_2(y)=x
\ee
i.e.
\be\label{eqminV1V2}
V'_2(V'_1(x))=x
\ee
Let us call $K$ the degree of that algebraic equation, i.e. the number of its solutions:
\be
(\ovl{x}_1,\ovl{y}_1),\dots,(\ovl{x}_K,\ovl{y}_K)
\ee
The eigenvalues of matrices $M_1$, $M_2$ which extremize $\tr (V_1(M_1)+V_2(M_2)-M_1M_2)$ must be among the $K$ solutions described above,
or can be trapped on contours stopping at hard edges.
The filling fractions $(\epsilon_1,\dots,\epsilon_{K+K_1+K_2})$ represent the number of eigenvalues equal to each solutions of \eq{eqminV1V2}, i.e.
there are $N\epsilon_1$ eigenvalues of $M_1$ equal to $\ovl{x}_1$,...
One must have:
\be
\sum_{i=1}^{K+K_1+K_2} \epsilon_i=1
\ee
The average number of eigenvalues of $M_1$ in the vicinity of a point $(\ovl{x},\ovl{y})$,
is a contour integral of the resolvent, along a contour which surrounds $(\ovl{x},\ovl{y})$:
\be\label{fixedfillfract}
2i\pi N \epsilon_i = -\oint_{{\cal A}_i} W(x)dx = \oint_{{\cal A}_i} Y(x) dx
\ee
and it can be non-zero only on irreducible cycles of the algebraic curve.

Thus, in the Formal matrix model, a set of irreducible cycle contour integrals are fixed.
One can verify that the number of filling fractions matches exactely the number of coefficients of $E(x,y)$ not determined by the pole behaviours,
Thus, \eq{fixedfillfract}  is sufficient to determine completely the rational function $E(x,y)$.

\section{Conclusion}

The purpose of this article was to write down the loop-equations for the so-called semiclassical 2-matrix model.
We find, that the loop equation becomes an algebraic equation $E(x,y)=0$ in the large $N$ limit, with poles at the poles of the potentials,
and at the hard edges.
The hard edges are such that the resolvent has a simple pole, and $dx$ has a zero, so that the differential form $ydx$ is regular.

The loop equation determines the form of the algebraic equation $E(x,y)$, but does not determine its coefficients.
The coefficients are determined by extra asumptions, related to which definition of the matrix model is consiered.
In the formal matrix model, the ${\cal A}$ cycle integrals of $ydx$ are fixed parameters of the model, and that determines $E(x,y)$ completely.
In the convergent matrix model, the real parts of both ${\cal A}$ and ${\cal B}$ cycle integrals of $ydx$ must vanish, and that determines $E(x,y)$ completely.

\medskip

Let us also remark that the function \eq{maineq} $E(x,y)$ is unchanged under the exchange $x\leftrightarrow y$, $V_1\leftrightarrow V_2$, $s\leftrightarrow \td{s}$,
which is the generalization of Matytsin's duality property \cite{matytsin}:
\be
X(Y(x))=x
\ee

\medskip

The consequences of that algebraic equation, can then be studied.
This is done for instance in \cite{Marcoratlfree}.
One can also expect to generalize the works of \cite{EOtrmixte}, or \cite{eynloop1mat}, or \cite{EOloop2mat}, i.e. the computation of all correlation functions and
their $1/N^2$ expansion,
and further, compute the expansion of the free energy \cite{eynm2m,eynm2mg1}.

\subsection*{Aknowledgements}
The author wants to thank Marco Bertola for numerous discussions, and for the motivation to compute those loop equations.
This work was partly supported by the european network ENIGMA (MRTN-CT-2004-5652).


\end{document}